\documentclass[twocolumn,prl,showpacs,superscriptaddress]{revtex4}
\usepackage{amsmath}
\usepackage{graphicx}
\usepackage{bm}
\usepackage{ulem}

\begin{document}

\title{Threshold for Chaos and Thermalization in One-Dimensional Mean-Field Bose-Hubbard model}

\author{Amy C. Cassidy}
\affiliation{Department of Physics and Astronomy, University of Southern
California, Los Angeles, CA 90089, USA}
\affiliation{Department of Physics, University of Massachusetts Boston, Boston MA 02125, USA}
\author{Douglas Mason}
\affiliation{Department of Physics, Harvard University, Cambridge, Massachusetts 02138, USA}
\author{Vanja Dunjko}
\affiliation{Department of Physics, University of Massachusetts Boston, Boston MA 02125, USA}
\author{Maxim Olshanii}
\affiliation{Department of Physics, University of Massachusetts Boston, Boston MA 02125, USA}

\date{\today}

\begin{abstract}
We study the threshold for chaos and its relation to thermalization in the 1D mean-field
Bose-Hubbard model, which in particular describes atoms in optical lattices.
We identify the threshold for chaos, which is finite in the thermodynamic limit, and show that it is indeed a precursor of thermalization.
Far above the threshold, the state of the system after relaxation is governed by the usual laws
of statistical mechanics.
\end{abstract}
\pacs{67.85.-d,03.75.Lm,05.45.Jn}
\maketitle

{\it Introduction}.--
We study the threshold for chaos and the ability to thermalize of
the 1D mean-field Bose-Hubbard model (BHM) \cite{zoller98}. The study of
thermalization in non-linear systems dates back to the early work of Fermi, Pasta, and Ulam (FPU) \cite{ulam55} on a non-linear string, modeled by anharmonically coupled oscillators.  It was expected that for a large number of degrees of freedom, even small nonlinearities would cause the system to thermalize, resulting in energy equipartition.  However, equipartition was not observed.
The absence of thermalization was eventually explained in two complementary ways: one in terms of closeness to an integrable system, the Korteweg-de Vries model \cite{zabusky65}, and another in terms of a chaos threshold
given by the theory of overlapping resonances put forth by Chirikov and Israilev \cite{chirikov59, chirikov66}.

Since then further studies on thermalization and approach to equilibrium
have been carried out in several classical field theories, including recent studies on the classical $\phi^4$ model
\cite{deVega04}, Nonlinear Klein-Gordon equation (NLKG) \cite{gerhardt02},
Non-Linear Schr\"{o}dinger equation (NLSE) \cite{lewenstein00,ablowitz89}, Discrete Non-Linear Schr\"{o}dinger equation (DNLS) \cite{ablowitz89,ablowitz93}
equivalent to BHM,
and Integrable Discrete Non-Linear Schr\"{o}dinger equation (IDNLS)\cite{ablowitz89}.

No conventional thermalization is expected in the NLSE and IDNLS, which are both integrable.
In NLKG, like in FPU, the ability of the system to reach thermal equilibrium in the course of time evolution
emerges only when the degree of nonlinearity exceeds a certain critical value (see \cite{chirikov66,livi85} for the thermalization threshold in FPU).
On the contrary, the $\phi^4$ model
eventually reaches equilibrium regardless of how small the nonlinearity is. In our paper we show that the BHM (along with the equivalent DNLS) belongs
to the former class.

Furthermore, we have compared two quantitative measures of thermalizability: maximal Lyapunov exponent
(whose positivity is a signature for chaos) and spectral entropy (which provides a distance to
thermal equilibrium). Both measures show a sharp threshold as one varies the nonlinearity strength, and the two
thresholds are undeniably close. Furthermore, we assert that the chaos threshold is governed only by the parameters and observables that are finite in the thermodynamic limit, and as a result it remains finite in that limit.

Our program is similar to a comprehensive comparison between
FPU and $\phi^4$ \cite{pettini90-91}, where, however, the existence of the
thermalization threshold in FPU is denied. 

In this paper we observe thermal behavior in time-averaged mean-field quantities.
Note that in recent work on thermalization in quantum systems, thermal properties emerge from individual quantum stationary states \cite{srednicki94,quantum-thermalization}.
Studies on the semi-classical regime suggest that the two are related, although open questions remain
\cite{srednicki94}.

Empirically, our system describes the motion of bosonic atoms in a one-dimensional tight-binding optical lattice \cite{zoller98,greiner01-esslinger04}.

{\it System of interest}.--
We study the mean-field dynamics of an interacting one-dimensional Bose gas on a lattice (1D Bose-Hubbard model (BHM))
with periodic boundary conditions.
The Hamiltonian in the momentum representation is
\begin{eqnarray}
&&H = \sum_{n} \left( \hbar\omega_n|\psi_n|^2 - \frac{\mu_{0}}{2} |\psi_n|^4 \right) +
\label{hamiltonian}
\\
&&\qquad\qquad \frac{\mu_0}{2}\sum_{\substack{i,j,n \\ n \ne i,j}} \psi_{i}^\ast \psi^\ast_{j} \psi^{}_{n} \psi^{}_{i+j-n}
\quad,
\nonumber
\end{eqnarray}
where the indices span the range $n,\,i,\,j = 0,\,\pm 1,\, \pm 2,\,\ldots,\,\pm \frac{N_{\mbox{\scriptsize s}}-1}{2}$
($N_{\mbox{\scriptsize s}}$ is supposed to be odd).
Throughout the text the wavefunction  $\psi_n$ is
normalized to unity:
$\sum_{n} |\psi_n|^2 = 1$.
The bare frequency of each momentum mode is given by
$\omega_n=-2J\cos\left(\frac{2\pi n}{N_{\mbox{\scriptsize s}}}\right)$,
and the coupling constant is $\mu_0 = U N_{\mbox{\scriptsize a}}/N_{\mbox{\scriptsize s}}$.
Here $J$ and $U$ are the nearest-neighbor site-hopping and on-site repulsion constants
of the standard Bose-Hubbard model, respectively, and $N_{\mbox{\scriptsize a}}$ is the number of atoms.
The canonical pairs are ${\cal Q}_{n} = \psi_{n}^{}, \, {\cal P}_{n} = i\hbar\psi_{n}^{\ast}$,
and the equations of motion are given by
$\frac{\partial}{\partial t}\psi_n = -\frac{i}{\hbar}\frac{\partial H}{\partial \psi_n^\ast}$.
We define the dimensionless non-linearity parameter, $\kappa$, to be the ratio between the typical interaction energy per site, $U(N_{\mbox{\scriptsize a}}/N_{\mbox{\scriptsize s}})^2$,  and the
hopping energy per site, $J N_{\mbox{\scriptsize a}}/N_{\mbox{\scriptsize s}}$:
\begin{equation}
\kappa \equiv \frac{\mu_{0}}{J} \equiv \frac{U(N_{\mbox{\scriptsize a}}/N_{\mbox{\scriptsize s}})}{J}
\quad.
\label{kappa}
\end{equation}
%
\textit{Chaos criterion and chaos threshold from Lyapunov exponents}.--
The standard signature of the chaotic nature of a region in phase space is that the separation between initially close trajectories grows exponentially with time,
for typical trajectories, as captured by a positive maximal Lyapunov exponent (MLE). In regular regions the separation
grows linearly \cite{chirikov79}, resulting in zero MLE. As we increase $\kappa$ in our system, we expect the phase space to change from being dominated
by regular regions for small $\kappa$ to being dominated by chaotic regions for large $\kappa$. In the present section, we use the MLEs to quantify this
transition to chaos, which, as we will see in the subsequent section, coincides with a relatively broad change from unthermalizability to complete thermalizability.

Consider two trajectories $\bm{x}(t)$ and $\widetilde{\bm{x}}(t)$ with initial points $\bm{x}_{0}$ and $\widetilde{\bm{x}}_{0}$, respectively. The separation
$\delta\bm{x}(t)=\widetilde{\bm{x}}(t)-\bm{x}(t)$ initially satisfies a linear differential equation, and the duration of this linear regime grows without bound as the initial separation
$\widetilde{\bm{x}}_{0}-\bm{x}_{0}$ goes to zero. The finite-time maximal Lyapunov exponent (FTMLE) corresponding to the phase-space point $\bm{x}_{0}$ \cite{eckhardt93-voglis94} is
\begin{eqnarray}
\lambda_{t_{\mbox{\scriptsize fin}}}(\bm{x}_{0})=
\lim_{\widetilde{\bm{x}}_{0}\to\bm{x}_{0}}\;
\frac{1}{t_{\mbox{\scriptsize fin}}}
\ln \frac{\|\widetilde{\bm{x}}(t_{\mbox{\scriptsize fin}})-\bm{x}(t_{\mbox{\scriptsize fin}})\|}{\|\widetilde{\bm{x}}_{0}-\bm{x}_{0}\|}
\,.
\label{lyapunov_exponent}
\end{eqnarray}
The limit $t_{\mbox{\scriptsize fin}}\to \infty$ gives the MLE, $\lambda_{\infty}(\bm{x}_{0})$. The FTMLEs are themselves of intrinsic interest and in the chaotic regime the average over the FTMLE converges to the standard MLE \cite{eckhardt93-voglis94,contopoulos78} .
We chose a convenient quantum
mechanical metric,
\mbox{$\|\widetilde{\bm{x}}-\bm{x}\|^{2} =$} $\sum_{n} |\widetilde{\psi}_{n}-\psi^{}_{n}|^2$ (see \cite{metric}).

Initially, we study the FTMLE on a 21-site lattice for a class of initial conditions where only the $k=0,\,\pm1$ modes are occupied. In this subspace we sample uniformly from the intersection of the
microcanonical shells in energy and norm; the energy is chosen to be the infinite temperature energy of the subsystem, and the norm is 1. For each value of $\kappa$, we sample 100 points, which we
set as the initial points $\bm{x}_{0}$. To each initial point we add a small random vector, as little as machine precision allows, to obtain the corresponding $\widetilde{\bm{x}}_{0}$'s. Each pair we
propagate for a time $t_{\mbox{\scriptsize fin}}$,  short enough to ensure linearity of the evolution of $\delta\bm{x}(t)$ but long enough
to be able to clearly
distinguish chaotic trajectories from regular ones on a plot of $\ln\delta\bm{x}(t)$ versus $t$:  the former
increase linearly, and the latter, logarithmically \cite{contopoulos78}. We also verify that the average of the FTMLE's over the ensemble of initial
conditions does not depend on $t_{\mbox{\scriptsize fin}}$ as long as both criteria above are satisfied.
\begin{figure}
\includegraphics[scale=.6]{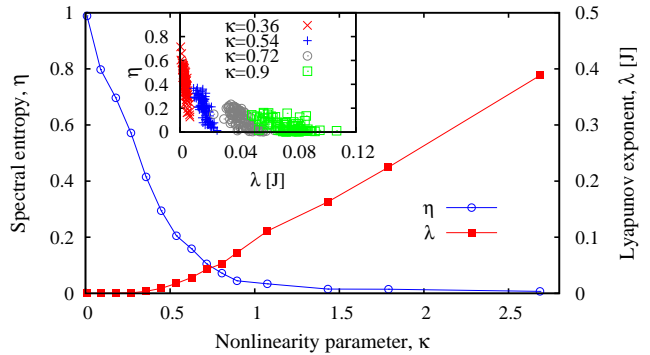}
\caption{\label{fig:lyapunov_exp} (color online). Averaged finite-time maximal Lyapunov exponent (FTMLE), $\lambda$,
and normalized spectral entropy, $\eta$, as functions of the nonlinearity, $\kappa$. $N_{\mbox{\scriptsize s}}$=21. Inset: Normalized spectral entropy of final time-averaged state versus FTMLE for
each of the 100 initial condition used to compute the averaged value for $\kappa=0.36$, $0.54$, $0.72$, $0.9$.}
\end{figure}
In Fig.~\ref{fig:lyapunov_exp} the averaged FTMLEs
are plotted as a function of the interaction strength.
There is a distinct regime with zero Lyapunov exponent for small $\kappa\lesssim 0.5$ and a strongly chaotic regime for $\kappa\gtrsim 1$ where all initial conditions have positive exponent.
%

{\it Thermalizability threshold from spectral entropy}.--
For coupled anharmonic oscillators, as in the FPU study, energy equipartition
among the normal momentum modes signified thermalization.
In the BHM, the additional conservation of the norm modifies the quantity that is equipartitioned.
To determine the best measure for the equipartition we use the variational Hartree-Fock Hamiltonian \cite{hartree-fock},
$H^{\mbox{\scriptsize HF}} = \sum_{n} \hbar\omega^{\mbox{\scriptsize HF}}_n|\psi_n|^2$,
where the set of Hartree-Fock energies
$\{\hbar\omega^{\mbox{\scriptsize HF}}_n\}$ was regarded as the variational field. This procedure gives
$\hbar\omega^{\mbox{\scriptsize HF}}_n = \hbar\omega_n + 2 \mu_0 N_{\mbox{\scriptsize a}} - \mu$,
where $\mu$ is the chemical potential.

The Hartree-Fock approximation is known to overestimate the interaction energy in the regime of strong interactions. For this reason,
we determine the temperature $T$ and the chemical potential using the time-averaged numerical
kinetic energy (along with the norm) instead of the total energy.
The temperature and the chemical potential were computed individually for each initial condition used.

The new quantity to be equipartitioned is the distribution of the Hartree-Fock energy,
$ q_n(t) = |\psi_n(t)|^2 \hbar\omega^{\mbox{\scriptsize HF}}_n
         /
         \sum_{n'} |\psi_{n'}(t)|^2 \hbar\omega^{\mbox{\scriptsize HF}}_{n'}
$.
A quantitative measure of the distance from thermodynamic equilibrium is the spectral
entropy $S(t) = -\sum_{n} q_n(t) \ln q_n(t)$,
or more conveniently the normalized spectral entropy \cite{livi85},
\begin{equation}
\eta(t)=\frac{S_{\text{max}} - S(t)}{S_{\text{max}} - S(0)},
\end{equation}
where $S_{\text{max}}= \ln N_{\mbox{\scriptsize s}}$ is the maximum entropy, which occurs for
complete equipartition of $q_n$.
In Fig.~\ref{fig:lyapunov_exp} the spectral entropy of the final time-averaged state, also averaged over 100 initial states
(drawn from the same ensemble that was used for the Lyapunov exponent calculation)
is plotted for each value of $\kappa$.
For large nonlinearities, $\kappa \gtrsim 1$, the normalized spectral entropy goes to zero, indicating
remarkable agreement between the final state and the thermal predictions. Note that this corresponds to the chaos threshold observed previously. Furthermore, we verified that for large $\kappa$, the fluctuations in kinetic energy scale as $\sqrt{N_{\mbox{\scriptsize s}}}$, confirming their thermal nature.
For $\kappa \lesssim .5$ the normalized spectral entropy is above $.5$ signifying that
during the time evolution the state of the system remains close to the initial state.
As seen in the inset of Fig.~\ref{fig:lyapunov_exp}, an individual initial state with larger FTMLE tends to have lower spectral entropy, i.e. to relax to a state which is closer to the thermal one.
Beginning at $\kappa \approx 0.5$, where the averaged FTMLE is substantially non-zero, some of the initial states thermalize completely.

\begin{figure}
\includegraphics[scale=.62]{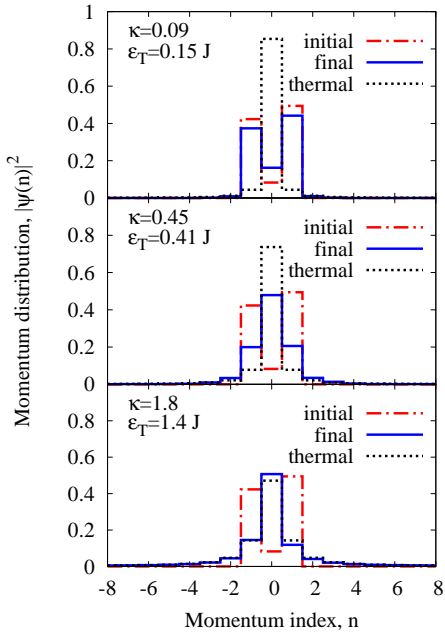}
\caption{\label{fig:avg_mom_dist} (color online). Initial, final and Hartree-Fock thermal momentum distributions for $\kappa=0.09,0.45,1.8,$ starting from the same initial state. N=21. The initial state is a representative state and the final state is time averaged. $\epsilon_T$ is the total energy per particle.
        }
\end{figure}

In Fig.~\ref{fig:avg_mom_dist}, the initial and time-averaged momentum distributions of
a representative state are plotted for $\kappa=0.09,0.36$ and $0.9$, along with the thermal Hartree-Fock predictions,
$\langle |\psi_n|^2 \rangle = (T/N_{\mbox{\scriptsize a}})/(\hbar \omega_n + 2 \mu_0 N_{\mbox{\scriptsize a}} - \mu)$.

{\it Chaos Threshold for Different Lattice Sizes}.--
Let us start from the notion that the parameter $\kappa$ introduced in (\ref{kappa}) is the only dimensionless combination of the parameters
of the problem that remain finite in the thermodynamic limit, $N_{\mbox{\scriptsize s}} \to \infty$, $N_{\mbox{\scriptsize a}}/N_{\mbox{\scriptsize s}} = \mbox{const}, J = \mbox{const},
U = \mbox{const}$. Curiously, the chaos threshold for $N_{\mbox{\scriptsize s}} = 21$  is at $\kappa \approx .5$, i.e. $\kappa \sim 1$.
Another observation comes from a related work \cite{lewenstein00} on chaos threshold in NLSE with hard-wall boundary conditions. The authors find that the boundary between regular and chaotic motions of momentum mode, $n$, is given by $(\mu_{0}|\psi_n|^2)/(\hbar\omega_{1} n) \sim 1$, where $\hbar\omega_{1}$ is the lowest
excitation energy, {\it e.g.} the energy of the first excited mode in the case of the Hamiltonian (\ref{hamiltonian}). Assuming that the shape of the
momentum distribution $|\psi_n|^2$ as a function of $n/N_{\mbox{\scriptsize s}}$ should be fixed in the thermodynamic limit, the left-hand-side of the above relationship also remains finite.
These observations lead to a conjecture that the chaos criterion involves only the intensive parameters and observables, {\it i.e. those that are finite in the thermodynamic limit}.

\begin{figure}
\includegraphics[scale=.65]{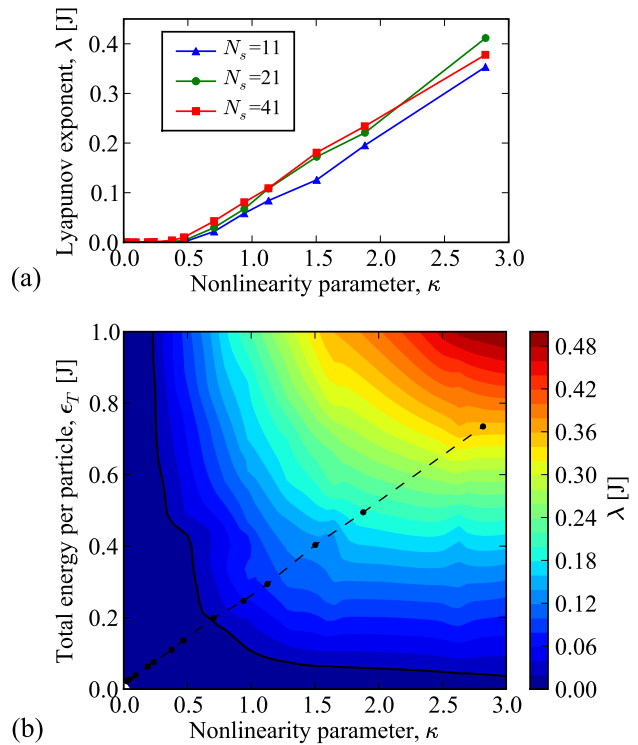}
\caption{\label{fig:lamda_size_scaling} (color online). (a) Averaged Finite Time Lyapunov exponent, $\lambda/J$, for three different system sizes, $N_{\mbox{\scriptsize s}}=11,21,41$.
For each $\kappa$, the same energy-per-particle was used for each lattice size. (b) Contour lines of the Lyapunov exponent versus the nonlinearity, $\kappa$ and energy-per-particle, $\epsilon_T = (H-H_{0})/N_{\mbox{\scriptsize a}}$, where $H$ is the Hamiltonian (\ref{hamiltonian}), and $H_{0} = -2J + (1/2)\mu_{0}$ is the ground state value of $H$. $N_{\mbox{\scriptsize s}}=11$.
The first contour line corresponds to $\lambda_c=0.02$. The circles and dotted line give the total energies (per particle) used in the calculation for (a).
        }
\end{figure}
Our test for the above conjecture is based on the fact that for a chaotic motion the majority of the trajectories cover the whole available phase space, and as a result the MLE becomes, for a given set of parameters, a function of just the conserved quantities: energy and norm. This implies that for the same energy-per-particle, norm, and nonlinearity
parameter $\kappa$, the Lyapunov exponents for different lattice sizes should be similar. In Fig.~\ref{fig:lamda_size_scaling}a the averaged FTMLE is plotted for three different
lattices, $N_{\mbox{\scriptsize s}}=11$, $21$, and 41. For each $\kappa$, the same energy-per-particle (in units of J) is used for all three lattices. The corresponding energies are shown by the solid line in Fig.~\ref{fig:lamda_size_scaling}b.
From the plot it is indeed evident that the averaged FTMLE is universal with respect to the size of the lattice and that the values for $N_{\mbox{\scriptsize s}}=11$ already give a very good estimate of both
the value of the averaged FTMLE and the threshold.

{\it Two Parametric Theory of the Chaos Threshold}.--
The universality observed above suggests the most relevant pair of variables for mapping the chaos threshold, namely $\kappa$ and the total energy-per-particle, $\epsilon_{T}/J$. (In FPU one variable is sufficient, ultimately because there is one less conserved quantity.)
In Fig.~\ref{fig:lamda_size_scaling}b contour lines of the averaged FTMLE for $N_{\mbox{\scriptsize s}}=11$ are plotted versus the nonlinearity parameter and energy-per-particle.
We use two sets of initial conditions with $n=0,\pm 1$ and $n=0,\pm 1,\pm 2$ momentum modes occupied.

One can observe a plateau in the averaged FTMLE for $\lambda \lesssim
\lambda_c=0.02$, given by the solid line. After crossing the critical
line the averaged FTMLE increases with uniform slope. The critical line resembles
a hyperbola with the point of closest approach to the
origin at $(\kappa,\,\epsilon_{T}) \sim (0.5,\,0.2 J)$, so that the hopping
parameter
$J$ appears to be a relevant energy scale. This is probably not an
accident: for $\epsilon_{T} \gg J$ the dispersion law $\omega_{n}$
begins to deviate from the (quadratic) dispersion law of the integrable
NLSE with periodic boundary conditions.

{\it Summary and outlook}.--
In this paper we consider the dynamics of atoms
in an optical lattice from the point of view of chaos theory. We identify
the threshold for chaos and show that it corresponds to the onset of thermalization.
Far above the threshold, the final state of the system is governed by the usual statistical mechanics.

We see two potential applications of our results. First, in quantum nonequilibrium dynamics, our results can serve as a guide for identifying the dominant effects preventing thermalization
in optical lattices. 
Based on the studies of the validity of
the classical field theory for Bose condensates \cite{castin04-00-kagan97} our results will
apply for the lattice site occupations satisfying
$N_{\mbox{\scriptsize a}}/N_{\mbox{\scriptsize s}} \gg \mbox{max}(\kappa,\, 1)\, \mbox{max}((\Delta n/N_{\mbox{\scriptsize s}})^{-1},\, 1)$, where $\Delta n$ is the typical width of the momentum distribution.
We note that the Mott regime, $ N_{\mbox{\scriptsize a}}= \mbox{integer}\times N_{\mbox{\scriptsize s}}$, $\Delta n = N_{\mbox{\scriptsize s}}$, $ U/J\geq 2.2\, N_{\mbox{\scriptsize a}}/N_{\mbox{\scriptsize s}}$ \cite{hamer79}, lies well outside of the above criteria.

Second, in chip-based atom interferometry with dense Bose condensates \cite{anderson05},  our results illustrate the fact that {\it nonlinear instabilities cannot affect the performance of interferometric schemes}. 
Recall that the force fields used in interferometry are usually periodic with a
period $L = \lambda/2$, where $\lambda = 2\pi/k$, and $k$ is the wavevector of light used
to generate the interferometric elements.
For spatially uniform initial conditions, the time evolution can be described by a NLSE with periodic
boundary conditions. In turn, the NLSE constitutes the continuum limit of our model, $N_{\mbox{\scriptsize s}} \to \infty$, where we keep constant
the ground-state chemical potential $\mu_{0}$, the size of system $L$, and the ratio between the energy-per-particle $E_{T}$ and
the so-called recoil energy $E_{\mbox{\scriptsize R}} \equiv \hbar^2 k^2/2m = \pi^2\,J/N_{\mbox{\scriptsize s}}$. In this limit both the parameter $\kappa$ and the $\epsilon_{T}/J$ ratio tend to zero as $N_{\mbox{\scriptsize s}}^{-2}$, i.e. towards the origin in Fig.~\ref{fig:lamda_size_scaling}(b), where the motion has no dynamical instabilities.

\begin{acknowledgments}
We are grateful to Bala Sundaram, Kurt Jacobs, Isabelle Bouchoule, Boris Svistunov, and Anatoli Polkovnikov for enlightening discussions on the subject. This work was supported by grants from the Office of Naval Research ({\it N00014-06-1-0455}) and the National Science Foundation ({\it PHY-0621703} and {\it PHY-0754942}).
\end{acknowledgments}

%

%
\end{document}